\documentclass[11pt]{article}
\usepackage{amstex,amssymb}

\newcommand{\qed}{\rule{3mm}{3mm}}

\newcommand{\ed}{{\bf 1}}

\newcommand{\bL}{{\bf L}}

\newcommand{\bS}{{\bf S}}

\newcommand{\bbL}{{\mathbb L}}
\newcommand{\bbS}{{\mathbb S}}

\newcommand{\cL}{{\cal L}}

\newcommand{\cU}{{\cal U}}

\newcommand{\gog}{{\mathfrak g}}

\newtheorem{theorem}{Theorem}

\begin{document}

\begin{center}
{\large\bf 
The motion of a rigid body in a quadratic potential:\\
an integrable discretization}\end{center}
\vspace{0.5cm}

\begin{center}
{\sc Yuri B.\,Suris}
\end{center}
\begin{center}
Fachbereich Mathematik, Technische Universit\"at Berlin, \\
Str. 17 Juni 136, 10623 Berlin, Germany\\  E--mail: 
{\tt suris} {\makeatother @@ \makeatletter}{\tt sfb288.math.tu-berlin.de}
\end{center}
\vspace{0.5cm}

\begin{abstract}
The motion of a rigid body in a quadratic potential is an important example
of an integrable Hamiltonian system on a dual to a semidirect product Lie 
algebra ${\rm so}(n)\ltimes {\rm Symm}(n)$. We give a Lagrangian derivation of 
the corresponding equations of 
motion, and introduce a discrete time analog of this system. The construction 
is based on the discrete time Lagrangian mechanics on Lie groups, accompanied 
with the discrete time Lagrangian reduction. The resulting multi--valued map
(correspondence) on the dual to ${\rm so}(n)\ltimes {\rm Symm}(n)$ is Poisson
with respect to the Lie--Poisson bracket, and is also completely
integrable. We find a Lax representation based on matrix factorisations,
in the spirit of Veselov--Moser.
\end{abstract}

\newpage

\section{Introduction}
The rigid body dynamics are rich with problems interesting from the 
mathematical point of view, in particular, with integrable problems.
Certainly, the most famous ones are the three integrable cases, named after
Euler, Lagrange, and Kovalevskaya, of the rotation of a heavy rigid 
body around a fixed point in a homogeneous gravity field. They were 
discovered in the 18th and the 19th century, and can be called classical. 
However, the list of intergrable problems of the rigid body dynamics 
is by far not exhausted by these ones.

In the present paper we turn our attention to the rotation of a rigid body
around its fixed center of mass in an arbitrary quadratic potential.
The integrability of this problem is a much more recent observation 
due to Reyman [R] and Bogoyavlensky [B]. (However, some particular case
of this result was given already by Brun [Br]; the equations of motion in
this case are identical with those describing the integrable case of the
motion of a rigid body in an ideal fluid, due to Clebsch [C]).

A problem we solve in the present paper, is a construction of an
{\it integrable discretization} of the above mentioned integrable
mechanical system. Our construction is close in spirit to the work 
by Moser and Veselov [MV], where they used an approach based on the
discrete time Lagrangian mechanics. Several important integrable mechanical
problems were discretized in [MV], including the Euler top and the
Neumann system. The general theory of discrete time Lagrangian mechanics
on Lie groups was developed further in [BS1], where the list of Moser
and Veselov was extended by a discrete time Lagrange top. A further
development of the general theory was undertaken in [BS2], where general
discrete time Euler--Poisson equations on semidirect product Lie algebras 
were obtained as a result of a redution procedure, applied to discrete
time Lagrangian systems on Lie groups. A discretization of a top in a
quadratic potential, achieved in the present paper, serves as a spectacular
illustration to the abstract constructions in [BS2].

We recall the general theory of the Lagrangian reduction in the continuous time
and in the discrete time contexts, respectively, in Sect. 2 and 3. Further,
we give in Sect.4 a Lagrangian derivation of the equations of motion of a 
rigid body in a quadratic potential. In Sect. 5 we introduce the discrete 
time analogs of these constructions. Finally, in Sect. 6 the specialization
of these results for the Clebsch case is given. Conclusions are contained in
Sect. 7.

\setcounter{equation}{0}
\section{Lagrangian mechanics and Lagrangian reduction on $TG$}

Recall that a continuos time Lagrangian system is defined by a smooth function
$\bL(g,\dot{g})\,:\,TG\mapsto{\mathbb R}$ on the tangent bundle of a smooth
manifold $G$. The function $\bL$ is called the {\it Lagrange function}. We
will be dealing here only with the case when $G$ carries an additional
structure of a {\it Lie grioup}. For an arbitrary function 
$g(t)\,:\,[t_0,t_1]\mapsto G$ one can consider the {\it action functional}
\begin{equation}\label{action}
\bS=\int_{t_0}^{t_1}\bL(g(t),\dot{g}(t))dt\;.
\end{equation} 
A standard argument shows that the functions $g(t)$ yielding extrema of this
functional (in the class of variations preserving $g(t_0)$ and $g(t_1)$), 
satisfy with necessity the {\it Euler--Lagrange equations}. In local 
coordinates $\{g^i\}$ on $G$ they read:
\begin{equation}\label{EL gen}
\frac{d}{dt}\left(\frac{\partial\bL}{\partial\dot{g}^i}\right)=
\frac{\partial\bL}{\partial g^i}\;.
\end{equation}
The action functional $\bS$ is independent of the choice of local coordinates,
and thus the Euler--Lagrange equations are actually coordinate independent as
well. For a coordinate--free description in the language of differential
geometry, see \cite{A}, \cite{MR}.

Introducing the quantities 
\begin{equation}\label{Pi}
\Pi=\nabla_{\dot{g}}\bL\in T_g^* G\;,
\end{equation}
one defines the {\it Legendre transformation}:
\begin{equation}\label{Legendre gen}
(g,\dot{g})\in TG\;\mapsto\; (g,\Pi)\in T^*G\;.
\end{equation}
If it is invertible, i.e. if $\dot{g}$ can be expressed through $(g,\Pi)$, 
then the Legendre transformation of the Euler--Lagrange equations 
(\ref{EL gen}) yield  a {\it Hamiltonian system} on $T^*G$ with respect to the 
standard symplectic structure on $T^*G$ and with the Hamilton function
\begin{equation}\label{Ham gen}
H(g,\Pi)=\langle \Pi,\dot{g}\rangle-\bL(g,\dot{g})\;,
\end{equation}
(where, of course, $\dot{g}$ has to be expressed through $(g,\Pi)$). 

When working with the tangent bundle of a Lie group, it is convenient 
to trivialize it, translating all vectors to the group unit by left or right 
multiplication. We consider only the left trivialization here:
\begin{equation}\label{left triv}
(g,\dot{g})\in TG\;\mapsto\;(g,\Omega)\in G\times\gog \;,
\end{equation}
where
\begin{equation}\label{Omega}
\Omega=L_{g^{-1}*}\dot{g} \quad\Leftrightarrow\quad \dot{g}=L_{g*}\Omega\;.
\end{equation}
Denote the Lagrange function pushed through the above map, by
$\bL^{(l)}(g,\Omega)\,:\, G\times\gog\mapsto{\mathbb R}$, so that  
\begin{equation}\label{Lagr left}
\bL^{(l)}(g,\Omega)=\bL(g,\dot{g})\;,
\end{equation}

The trivialization (\ref{left triv}) of the tangent bundle $TG$ induces
the following trivialization of the cotangent bundle $T^*G$:
\begin{equation}\label{left triv *}
(g,\Pi)\in T^*G\;\mapsto\;(g,M)\in G\times\gog^* \;,
\end{equation}
where 
\begin{equation}\label{M}
M=L_g^*\Pi \quad\Leftrightarrow\quad \Pi=L_{g^{-1}}^* M \;.
\end{equation}

We now consider the Lagrangian reduction procedure, in the case when the
Lagrange function is symmetric with respect to the left action of a certain
subgroup of $G$, namely an isotropy subgroup of some element in the
representation space of $G$. So, the next ingredient of our construction is a 
representation $\Phi: G\times V\mapsto V$ of a Lie group $G$ in a linear 
space $V$; we denote it by
\[
\Phi(g)\cdot v \quad\text{for} \quad g\in G\;,\;\; v\in V\;.
\]
We denote also by $\phi$ the corresponding representation of the Lie algebra 
$\gog$ in $V$:
\begin{equation}
\phi(\xi)\cdot v=\left.\frac{d}{d\epsilon}\Big(\Phi(e^{\epsilon\xi})\cdot
v\Big)\right|_{\epsilon=0} \quad \text{for} \quad \xi\in\gog\;, \;\; v\in V\;.
\end{equation}
The map $\phi^*:\gog\times V^*\mapsto V^*$  defined by
\begin{equation}\label{phi star}
\langle \phi^*(\xi)\cdot y,v\rangle=\langle y,\phi(\xi)\cdot v\rangle\qquad
\forall v\in V\;,\; y\in V^*\;,\;\xi\in\gog\;,
\end{equation}
is an anti--representation of the Lie algebra $\gog$ in $V^*$.
We shall use also the bilinear operation 
$\diamond\,: V^*\times V\mapsto \gog^*$ introduced in [HMR, CHMR] and defined as
follows: let $v\in V$, $y\in V^*$, then 
\begin{equation}\label{diamond op}
\langle y\diamond v, \xi\rangle=-\langle y,\phi(\xi)\cdot v\rangle\qquad
\forall \xi\in\gog\;.
\end{equation}
(Notice that the pairings on the left--hand side and on the right--hand side 
of the latter equation are defined on different spaces).
\vspace{2mm}

Fix an element $a\in V$, and consider the isotropy subgroup $G^{[a]}$ of $a$, 
i.e.
\begin{equation}\label{G[a]}
G^{[a]}=\{h\,:\;\Phi(h)\cdot a=a\}\subset G \;.
\end{equation}  
Suppose that the Lagrange function $\bL(g,\dot{g})$ is invariant under 
the action of $G^{[a]}$ on $TG$ induced by left translations on $G$:
\begin{equation}\label{left action}
\bL(hg,L_{h*}\dot{g})=\bL(g,\dot{g})\;, \quad h\in G^{[a]}\;.
\end{equation}
The corresponding invariance property of $\bL^{(l)}(g,\Omega)$ is expressed as:
\begin{equation}\label{left action for red}
\bL^{(l)}(hg,\Omega)=\bL^{(l)}(g,\Omega)\;, \quad h\in G^{[a]}\;.
\end{equation}
We want to reduce the Euler--Lagrange equations with respect to this left 
action. As a section $(G\times\gog)/G^{[a]}$ we choose the set 
$\gog\times O_{a}$, where $O_{a}$ is the orbit of $a$ under the action $\Phi$:
\begin{equation}\label{orbit}
O_a=\{\Phi(g)\cdot a\,,\;g\in G\}\subset V\;.
\end{equation} 
The reduction map is
\begin{equation}
(g,\Omega)\in G\times\gog\;\mapsto\; (\Omega,P)\in \gog\times O_a\;, \qquad 
\text{where}  \qquad P=\Phi(g^{-1})\cdot a\;,
\end{equation}
so that the reduced Lagrange function $\cL^{(l)}\,:\,\gog\times O_a
\mapsto{\mathbb R}$ is defined as 
\begin{equation}\label{left red Lagr}
\cL^{(l)}(\Omega,P)=\bL^{(l)}(g,\Omega)\;,\quad{\rm where}\quad
 P=\Phi(g^{-1})\cdot a\;.
\end{equation}
The reduced Lagrangian $\cL^{(l)}(\Omega,P)$ is well defined, because from
\[
P=\Phi(g_1^{-1})\cdot a=\Phi(g_2^{-1})\cdot a
\]
there follows $\Phi(g_2g_1^{-1})\cdot a=a$, so that $g_2g_1^{-1}\in G^{[a]}$, 
and $\bL^{(l)}(g_1,\Omega)=\bL^{(l)}(g_2,\Omega)$.
 
\begin{theorem} {\rm [HMR, CHMR]}

{\rm a)} Under the left trivialization $(g,\dot{g})\mapsto(g,\Omega)$ and the
subsequent reduction $(g,\Omega)\mapsto(\Omega,P)$, the Euler--Lagrange 
equations {\rm (\ref{EL gen})} become the following {\bf Euler--Poincar\'e 
equations}:
\begin{equation}\label{EL left Ham red}
\left\{\begin{array}{l}
\dot{M}={\rm ad}^*\,\Omega\cdot M+\nabla_P\cL^{(l)}\diamond P\;,\\ \\
\dot{P}=-\phi(\Omega)\cdot P\;,\end{array}\right.
\end{equation}
where
\begin{equation}\label{M thru L red}
M=L_g^*\Pi=\nabla_{\Omega}\cL^{(l)}\;.
\end{equation}

{\rm b)} If the ``Legendre transformation''
\begin{equation}\label{Legendre left red}
(\Omega,P)\in \gog\times O_a\mapsto (M,P)\in \gog^*\times O_a\;,
\end{equation}
is invertible, then {\rm(\ref{EL left Ham red})} is a Hamiltonian system 
of  $\gog^*\times O_a$ with the Hamilton function
\[
H(M,P)=\langle M,\Omega\rangle-\cL^{(l)}(\Omega,P)\;,
\]
with respect to the Poisson bracket given by
\begin{equation}\label{PB left red}
\{F_1,F_2\}=\langle M,[\nabla_M F_1,\nabla_M F_2]\,\rangle+
\langle \nabla_P F_1,\phi(\nabla_M F_2)\cdot P\,\rangle-
\langle \nabla_P F_2,\phi(\nabla_M F_1)\cdot P\,\rangle
\end{equation}
for two arbitrary functions 
$F_{1,2}(M,P):\gog^*\times O_a\mapsto{\mathbb R}$. 
\end{theorem}

{\bf Remark 1.} The formula (\ref{PB left red}) defines a Poisson bracket 
not only on $\gog^*\times O_a$, but on all of $\gog^*\times V$. Rewriting
this formula as 
\begin{equation}\label{PB left red once more}
\{F_1,F_2\}=\langle M,[\nabla_M F_1,\nabla_M F_2]\,\rangle+
\langle  P, \phi^*(\nabla_M F_2)\cdot \nabla_P F_1-
\phi^*(\nabla_M F_1)\cdot \nabla_P F_2\,\rangle
\end{equation}
one immediately identifies this bracket with the Lie--Poisson bracket of the 
semiproduct Lie algebra $\gog\ltimes V^*$ corresponding to the representation
$-\phi^*$ of $\gog$ in $V^*$.
\vspace{2mm}

{\bf Remark 2.} In an important particular case of constructions of this 
section, the vector space is chosen as the Lie algebra 
of our basic Lie group: $V=\gog$, the group representation is the adjoint one:
$\Phi(g)\cdot v={\rm Ad}\; g\cdot v$, so that $\phi(\xi)\cdot v={\rm ad\;}\xi
\cdot v=[\xi,v]$, and the bilinear operation $\diamond$ is nothing but
the coadjoint action of $\gog$ on $\gog^*$: 
$y\diamond v={\rm ad}^*\, v\cdot y$. This is the framework, e.g., for the
heavy top mechanics.

\setcounter{equation}{0}
\section{Lagrangian mechanics and Lagrangian reduction on $G\times G$}

We now turn to the discrete time analog of these constructions,
Our presentation of the general discrete time Lagrangian mechanics
is an adaptation of the Moser--Veselov construction [V], [MV] for the case 
when the basic manifold is a Lie group. The presentation of the discrete time
Lagrangian reduction follows [BS1], [BS2]. Almost all constructions and results 
of the continuous time Lagrangian mechanics have their discrete time analogs. 
The only exception is the existence of the ``energy'' integral (\ref{Ham gen}).

Let $\bbL(g,\hat{g})\,:\,G\times G\mapsto{\mathbb R}$ be a smooth function, 
called the (discrete time) {\it Lagrange function}. For an arbitrary sequence 
$\{g_k\in G,\;k=k_0,k_0+1,\ldots,k_1\}$ one can consider the {\it action 
functional}
\begin{equation}\label{dS}
\bbS=\sum_{k=k_0}^{k_1-1}\bbL(g_k,g_{k+1})\;.
\end{equation}
Obviously, the sequences $\{g_k\}$ delivering extrema of this functional
(in the class of variations preserving $g_{k_0}$ and $g_{k_1}$), satisfy
with necessity the {\it discrete Euler--Lagrange equations}:
\footnote{For the notations from the Lie groups theory used in this and 
subsequent sections see, e.g., [BS1]. In particular, for an arbitrary smooth 
function $f:G\mapsto{\mathbb R}$ its right Lie derivative $d\,'f$ and left Lie 
derivative $df$ are functions from $G$ into $\gog^*$ defined via the formulas
\[
\langle df(g),\eta\rangle=\left.\frac{d}{d\epsilon}\,f(e^{\epsilon\eta}g)
\right|_{\epsilon=0}\;,\qquad 
\langle d\,'f(g),\eta\rangle=\left.\frac{d}{d\epsilon}\,f(ge^{\epsilon\eta})
\right|_{\epsilon=0}\;,\qquad \forall \eta\in\gog\;,
\]
and the gradient $\nabla f(g)\in T^*_gG$ is defined as
\[
\nabla f(g)=R^*_{g^{-1}}\,df(g)=L^*_{g^{-1}}\,d\,'f(g).
\]
}
\begin{equation}\label{dEL}
\nabla_1\bbL(g_k,g_{k+1})+\nabla_2\bbL(g_{k-1},g_k)=0\;.
\end{equation}
Here $\nabla_1\bbL(g,\hat{g})$ ($\nabla_2\bbL(g,\hat{g})$) denotes the 
gradient of $\bbL(g,\hat{g})$ with respect to the first argument $g$ 
(resp. the second argument $\hat{g}$). So, in our case, when $G$ is a Lie 
group and not just a general smooth manifold, the equation (\ref{dEL}) is 
written in a coordinate free form, using the intrinsic notions of the Lie 
theory. As pointed out above, an invariant formulation of the Euler--Lagrange 
equations in the continuous time case is more sophisticated. This seems 
to underline the fundamental character of discrete Euler--Lagrange equations. 

The equation (\ref{dEL}) is an implicit equation for $g_{k+1}$. In 
general, it has more than one solution, and therefore defines a correspondence
(multi--valued map)  $(g_{k-1},g_k)\mapsto(g_k,g_{k+1})$. To discuss symplectic
properties of this correspondence, one defines:
\begin{equation}\label{dPi}
\Pi_k=\nabla_2\bbL(g_{k-1},g_k)\in T^*_{g_k}G\;.
\end{equation} 
Then (\ref{dEL}) may be rewritten as the following system:
\begin{equation}\label{dEL syst}
\left\{\begin{array}{l}
\Pi_k=-\nabla_1\bbL(g_k,g_{k+1}) \\ \\ \Pi_{k+1}=\nabla_2\bbL(g_k,g_{k+1}) 
\end{array}\right.
\end{equation}
This system defines a (multivalued) map $(g_k,\Pi_k)\mapsto(g_{k+1},\Pi_{k+1})$
of $T^*G$ into itself. More precisely, the first equation in (\ref{dEL syst})
is an implicit equation for $g_{k+1}$, while the second one allows for the
explicit and unique calculation of $\Pi_{k+1}$, knowing $g_k$ and $g_{k+1}$. As 
demonstrated in \cite{V}, \cite{MV}, this map $T^*G\mapsto T^*G$ is symplectic
with respect to the standard symplectic structure on $T^*G$.

The tangent bundle $TG$ does not appear in the discrete time context at all. 
On the contrary, the cotangent bundle $T^*G$ still plays an important role in 
the discrete time theory, as the phase space with the canonical invariant 
symplectic structure. The left trivialization of $T^*G$ is same as in the
continuous time case: 
\begin{equation}\label{d left triv *}
(g_k,\Pi_k)\in T^*G \;\mapsto\; (g_k,M_k)\in G\times\gog^*\;,
\end{equation}
where 
\begin{equation}\label{d M}
M_k=L_{g_k}^*\Pi_k \quad\Leftrightarrow\quad \Pi_k=L_{g_k^{-1}}^* M_k\;.
\end{equation}
Consider also the map
\begin{equation}\label{d left triv}
(g_k,g_{k+1})\in G\times G \;\mapsto\; (g_k,W_k)\in G\times G\;,
\end{equation}
where
\begin{equation}\label{W}
W_k=g_k^{-1}g_{k+1} \quad\Leftrightarrow\quad g_{k+1}=g_kW_k\;.
\end{equation}
Denote the Lagrange function pushed through (\ref{d left triv}) by
\begin{equation}\label{dLagr left}
\bbL^{(l)}(g_k,W_k)=\bbL(g_k,g_{k+1})\;.
\end{equation}

Suppose that the Lagrange function $\bbL(g,\hat{g})$ is invariant under 
the action of $G^{[a]}$ on $G\times G$ induced by left translations on $G$:
\begin{equation}\label{d left action}
\bbL(hg,h\hat{g})=\bbL(g,\hat{g})\;, \quad h\in G^{[a]}\;.
\end{equation}
The corresponding invariance property of $\bbL^{(l)}(g,W)$ is expressed as:
\begin{equation}\label{d left action for red}
\bbL^{(l)}(hg,W)=\bbL^{(l)}(g,W)\;, \quad h\in G^{[a]}\;.
\end{equation}
We want to reduce the Euler--Lagrange equations with respect to this left 
action. As a section $(G\times G)/G^{[a]}$ we choose the set $G\times O_{a}$. 
The reduction map is
\begin{equation}
(g,W)\in G\times G\;\mapsto\; (W,P)\in G\times O_a\;, \qquad \text{where}
\qquad P=\Phi(g^{-1})\cdot a\;,
\end{equation}
so that the reduced Lagrange function $\Lambda^{(l)}\,:\,G\times O_a
\mapsto{\mathbb R}$ is defined as 
\begin{equation}\label{left red dLagr}
\Lambda^{(l)}(W,P)=\bbL^{(l)}(g,W)\;,\quad{\rm where}\quad
 P=\Phi(g^{-1})\cdot a\;.
\end{equation}
 
\begin{theorem} {\rm [BS1], [BS2]}

{\rm a)} Under the left trivialization $(g,\hat{g})\mapsto(g,W)$ and the
subsequent reduction $(g,W)\mapsto(W,P)$, the Euler--Lagrange 
equations {\rm (\ref{dEL})} become the following {\bf discrete 
Euler--Poincar\'e equations}:
\begin{equation}\label{dEL left Ham red}
\left\{\begin{array}{l}
{\rm Ad}^*\,W_k^{-1}\cdot M_{k+1}=M_k+\nabla_P
\Lambda^{(l)}(W_k,P_k)\diamond P_k\;,\\ \\
P_{k+1}=\Phi(W_k^{-1})\cdot P_k\;,\end{array}\right.
\end{equation}
where
\begin{equation}\label{d M thru L red}
M_k=d\,'_{\!W}\Lambda^{(l)}(W_{k-1},P_{k-1})\in\gog^*\;.
\end{equation}

{\rm b)} If the ``Legendre transformation''
\begin{equation}\label{dLegendre left red}
(W_{k-1},P_{k-1})\in G\times O_a\mapsto 
(M_k,P_k)\in \gog^*\times O_a\;,
\end{equation}
where $P_k=\Phi(W_{k-1}^{-1})\cdot P_{k-1}$, is invertible, then
{\rm(\ref{dEL left Ham red})} define a map $(M_k,P_k)\mapsto(M_{k+1},P_{k+1})$ 
of  $\gog^*\times O_a$ which is Poisson with respect to the Poisson 
bracket {\rm(\ref{PB left red})}.
\end{theorem}

The relation between the continuous time and the discrete time equations 
is established, if we set
\[
g_k=g\;,\qquad g_{k+1}=g+\varepsilon\dot{g}+O(\varepsilon^2)\;,\qquad
\bbL(g_k,g_{k+1})=\varepsilon\bL(g,\dot{g})+O(\varepsilon^2)\;;
\]
\[
P_k=P\;,\qquad W_k=\ed+\varepsilon\Omega+O(\varepsilon^2)\;,\qquad
\Lambda^{(l)}(W_k,P_k)=\varepsilon\cL^{(l)}(\Omega,P)+O(\varepsilon^2)\;.
\]

\setcounter{equation}{0}
\section{A rigid body in a quadratic potential}
The basic Lie group relevant for our main example is
\[
G={\rm SO}(n)\;,\quad {\rm so\;\;that} \quad \gog={\rm so}(n)
\]
(the ``physical'' rigid body corresponds to $n=3$). The scalar product
on $\gog$ is defined as
\[
\langle \xi,\eta\rangle=-\frac{1}{2}\,{\rm tr}(\xi\eta)\;, \quad
\xi,\eta\in\gog\;.
\]
This scalar product is used also to identify $\gog^*$ with $\gog$, so that
the previous formula can be considered also a pairing  betwenn the elements
$\xi\in\gog$ and $\eta\in\gog^*$. 

The group $G$ is a natural configuration space for problems related to
the rotation of a rigid body. Indeed, if $E(t)=\Big(e_1(t),\ldots,e_n(t)\Big)$ 
stands for the time evolution of a certain orthonormal frame firmly 
attached to the rigid body (so that all $e_k\in{\mathbb R}^n$), then
\[
E(t)=g^{-1}(t)E(0)\;\;\Leftrightarrow\;\; e_k(t)=g^{-1}(t)e_k(0)\quad\;
(1\le k\le n)\;,
\]
with some $g(t)\in G$. The Lagrange function of an arbitrary rigid body 
rotating abot a fixed point $0\in {\mathbb R}^n$ in a field with a quadratic 
potential
\begin{equation}\label{TQ pot}
\varphi(x)=\frac{1}{2}\sum_{i,j=1}^n a_{ij}x_ix_j\;,
\end{equation}
is equal to (cf. [B], [RSTS]):
\begin{equation}\label{TQ lagr 1}
\bL(g,\dot{g})=-\frac{1}{2}\,{\rm tr}(\Omega J\Omega)+\frac{1}{2}\,{\rm tr}
(gJg^{\rm T}A)\;,
\end{equation}
where 
\begin{itemize}
\item $\Omega=g^{-1}\dot{g}=g^{\rm T}\dot{g}$ is the angular velocity of the
rigid body in the body frame $E(t)$;
\item $J$ is a symmetric matrix (tensor of inertia of the rigid body); choosing
the frame $E(0)$ properly, we can assure this matrix to be {\it diagonal}, 
$J={\rm diag}(J_1,\ldots,J_n)$, which will be supposed from now on;
\item $A=(a_{ij})_{i,j=1}^n$ is a symmetric matrix of coefficients of the 
quadratic form $\varphi(x)$.
\end{itemize}
To include this Lagrange function in the framework of Sect. 2, we make the
following identifications:
\begin{itemize}
\item $V={\rm Symm}(n)$, the linear space of all $n\times n$ symmetric
matrices; we identify $V^*$ with $V$ via the following scalar product
on $V$:
\[
\langle v_1,v_2\rangle=\frac{1}{2}\,{\rm tr}(v_1v_2)\;, \;\; v_1,v_2\in V\;.
\]
\item The representation $\Phi$ of $G$ in $V$ is defined as
\[
\Phi(g)\cdot v=gvg^{-1}=gvg^{\rm T}\;\; {\rm for}\;\; g\in G, \;v\in V\;.
\]
\item Therefore the representation $\phi$ of $\gog$ in $V$ is given by
\[
\phi(\xi)\cdot v=[\xi,v]\;\;{\rm for}\;\; \xi\in\gog, \;v\in V\;,
\]
while the anti--representation $\phi^*$ of $\gog$ in $V^*$ is given by
\[
\phi^*(\xi)\cdot y=-[\xi,y]\;\;{\rm for}\;\; \xi\in\gog, \;y\in V^*\;,
\]
\item Finally, the bilinear operation $\diamond:\,V^*\times V\mapsto \gog^*$
is given by
\[
y \diamond v=-[y,v]\;\; {\rm for}\;\; y\in V^*\;,\;v\in V\;.
\]
\end{itemize}
Denoting now $P=g^{\rm T}Ag=\Phi(g^{-1})\cdot A$, we represent 
(\ref{TQ lagr 1}) in the form 
\begin{equation}\label{TQ lagr 2}
\bL(g,\dot{g})=\cL^{(l)}(\Omega,P)=
-\frac{1}{2}\,{\rm tr}(\Omega J\Omega)+\frac{1}{2}\,{\rm tr}(JP)=
-\frac{1}{2}\,{\rm tr}(\Omega J\Omega)+\langle J, P\rangle\;,
\end{equation}
which is manifestly invariant under the left action of the isotropy 
subgroup $G^{[A]}$. Now Theorem 1 is applicable, which delivers the
following equations of motion:
\begin{equation}\label{TQ}
\left\{\begin{array}{l} \dot{M}=[M,\Omega]+[P,J]\;,\\ \\ 
\dot{P}=[P,\Omega]\;,\end{array}\right.
\end{equation}
where
\begin{equation}\label{TQ M}
M=\nabla_{\Omega}\cL^{(l)}=J\Omega+\Omega J\quad \Leftrightarrow\quad
M_{jk}=(J_j+J_k)\Omega_{jk}\;.
\end{equation}
According to the general theory, the system (\ref{TQ}) is Hamiltonian
on the dual of the semidirect product Lie algebra $\gog\ltimes V^*$, with the
Hamilton function
\begin{equation}\label{TQ Ham}
H(M,P)=\frac{1}{2}\langle M,\Omega\rangle-\langle J, P\rangle=
\frac{1}{2}\sum_{j<k} \frac{M_{jk}^2}{J_j+J_k}-\frac{1}{2}\sum_{k=1}^n
J_kP_{kk}\;.
\end{equation}
Generic orbits in this Poisson phase space have dimension $n^2-n$ (the 
dimension of $\gog\times V^*$ is equal to $n(n-1)/2+n(n+1)/2=n^2$; the $n$ 
spectral invariants of $P\in V$ are Casimir functions of the Poisson bracket). 
In particular, for the ``physical'' case $n=3$ the dimension of the generic
orbit is equal to 6. Therefore the number of independent involutive integrals
of motion necessary for complete integrability is equal to $n(n-1)/2$
(equal to 3 for $n=3$).

A key observation of [R], [B] consists in the following Lax representation
of the above system:
\[
\dot{L}(\lambda)=[L(\lambda),B(\lambda)]\;,
\]
where
\[
L(\lambda)=P+\lambda M+\lambda^2 J^2\;,\qquad B(\lambda)=\Omega+\lambda J\;.
\]
(The key point in the straightforward verification of this statement is
the identity
\[
[M,J]+[J^2,\Omega]=0\;,
\]
which follows directly from (\ref{TQ M})).
The spectral invariants of the matrix $L(\lambda)$ provide us with the
necessary number of independent integrals of motion [B], and their involutivity
follows from the general $r$--matrix theory (cf. [R], [RSTS]).

\setcounter{equation}{0}
\section{A discrete time analog of a top in a quadratic potential}

To find a discrete analog of the Lagrange function (\ref{TQ lagr 1}), we
rewrite the latter once more as
\begin{equation}\label{TQ lagr 3}
\bL(g,\dot{g})=
\frac{1}{2}\,{\rm tr}(\dot{g}J\dot{g}^{\rm T})+
\frac{1}{2}\,{\rm tr}(gJg^{\rm T}A)\;.
\end{equation}
Let us introduce the following discrete analog:
\begin{equation}\label{dTQ lagr 1}
\bbL(g_k,g_{k+1})=
\frac{1}{2\epsilon}\,{\rm tr}\Big((g_{k+1}-g_k)J(g_{k+1}-g_k)^{\rm T}\Big)
+\frac{\epsilon}{2}\,{\rm tr}(g_{k+1}Jg_k^{\rm T}A)\;.
\end{equation}
The powers of $\epsilon$ are introduced in a way assuring the correct
asymptotics of the discrete time Lagrange function, namely
$\bbL(g_k,g_{k+1})\approx \epsilon\bL(g,\dot{g})$, as $g_k=g$ and 
$g_{k+1}\approx g+\epsilon\dot{g}$ (see the end of Sect. 3). Up to a constant, 
the function (\ref{dTQ lagr 1}) may be rewritten as
\begin{equation}\label{dTQ lagr 2}
\bbL(g_k,g_{k+1})=
-\frac{1}{\epsilon}\,{\rm tr}(g_{k+1}Jg_k^{\rm T})
+\frac{\epsilon}{2}\,{\rm tr}(g_{k+1}Jg_k^{\rm T}A)\;.
\end{equation}
This is representable also in terms of $W_k=g_k^{\rm T}g_{k+1}\in G$ and
$P_k=g_k^{\rm T}Ag_k\in O_{A}\subset V$:
\begin{equation}\label{dTQ lagr 3}
\bbL(g_k,g_{k+1})=\Lambda^{(l)}(W_k,P_k)=
-\frac{1}{\epsilon}\,{\rm tr}(W_kJ)
+\frac{\epsilon}{2}\,{\rm tr}(W_kJP_k)\;.
\end{equation}
(Recall that in the continuous limit one has $W_k\approx I+\epsilon\Omega$, so 
that $W_k^{\rm T}\approx I-\epsilon\Omega$).
\begin{theorem} The discrete time Euler--Lagrange equations for the 
Lagrange function {\rm(\ref{dTQ lagr 3})} are equivalent to the following
system:
\begin{equation}\label{dTQ}
\left\{\begin{array}{l}
M_k=\displaystyle\frac{1}{\epsilon}(W_kJ-JW_k^{\rm T})-
    \displaystyle\frac{\epsilon}{2}(P_kW_kJ-JW_k^{\rm T}P_k)\;\\ \\
M_{k+1}=\displaystyle\frac{1}{\epsilon}(JW_k-W_k^{\rm T}J)-
        \displaystyle\frac{\epsilon}{2}(JP_kW_k-W_k^{\rm T}P_kJ)\;,\\ \\
P_{k+1}=W_k^{\rm T}P_kW_k\;.
\end{array}\right.
\end{equation}
The multi--valued map (correspondence) $(M_k,P_k)\mapsto(M_{k+1},P_{k+1})$
described by {\rm(\ref{dTQ})} is Poisson with respect to the Lie--Poisson
bracket of the semidirect product Lie algebra $\gog\ltimes V^*$, where
$\gog={\rm so}(n)$, $V^*={\rm Symm}(n)$, and the representation $-\phi^*$
of $\gog$ in $V^*$ is defined as $-\phi^*(\xi)\cdot y=[\xi,y]$ for 
$\xi\in\gog$, $y\in V^*$.
\end{theorem}
{\bf Proof.} We are in a position to apply Theorem 2. To this end we first 
calculate
\[
\nabla_P\Lambda^{(l)}(W_k,P_k)=\frac{\epsilon}{2}(W_kJ+JW_k^{\rm T}),
\]
so that the equations of motion read:
\begin{equation}\label{dTQ 1}
\left\{\begin{array}{l}
W_kM_{k+1}W_k^{\rm T}=M_k+\displaystyle\frac{\epsilon}{2}\,
[P_k,W_kJ+JW_k^{\rm T}]\;\\ \\
P_{k+1}=W_k^{\rm T}P_kW_k\;.
\end{array}\right.
\end{equation}
These equations are, obviously, a discrete time approximation of 
(\ref{TQ}). They have to be supplemented by a discrete version of
(\ref{TQ M}), which reads:
\begin{equation}\label{dTQ M}
M_{k+1}=d\,'_{\!W}\Lambda^{(l)}(W_k,P_k)=\frac{1}{\epsilon}
(JW_k-W_k^{\rm T}J)-\frac{\epsilon}{2}(JP_kW_k-W_k^{\rm T}P_kJ)\;.
\end{equation}
Plugging this into the first equation of the system (\ref{dTQ 1}), we 
put this system into the form (\ref{dTQ}). \qed
\vspace{2mm}

The definition of the above correspondence (\ref{dTQ}) crucially depends
on the solvability of the first equation in (\ref{dTQ}) for $W_k\in G$.
The best approach to this problem, as well as to the integrability of the 
correspondence, is through the matrix factorizations. The following argument
is a generalization of the Moser--Veselov approach [MV] to the discrete time 
motion of the free $n$--dimensional rigid body.
 
Let us introduce the following matrices, depending on the spectral
parameter $\lambda$:
\begin{equation}\label{U}
\cU_k(\lambda)=\Big(I-\frac{\epsilon^2}{2}\,P_k\Big)W_k-\epsilon\lambda J\;,
\end{equation}
so that
\begin{equation}\label{U*}
\cU_k^{\rm T}(-\lambda)=W_k^{\rm T}\Big(I-\frac{\epsilon^2}{2}\,P_k\Big)
+\epsilon\lambda J\;.
\end{equation}
Denote also
\begin{equation}\label{dL}
L_k(\lambda)=\Big(I-\frac{\epsilon^2}{2}\,P_k\Big)^2-\epsilon^2\lambda M_k
-\epsilon^2\lambda^2 J^2=I-\epsilon^2\Big(P_k+\lambda M_k+\lambda^2J^2\Big)+
\frac{\epsilon^4}{4}\,P_k^2\;.
\end{equation}
Then a direct calculation allows us to verify the following statement.
\begin{theorem}
The equations of motion {\rm(\ref{dTQ})} are equivalent to the following
matrix factorizations:
\begin{equation}\label{dTQ fact}
\left\{\begin{array}{l}
L_k(\lambda)=\cU_k(\lambda)\,\cU_k^{\rm T}(-\lambda)\;,\\ \\
L_{k+1}(\lambda)=\cU_k^{\rm T}(-\lambda)\,\cU_k(\lambda)\;.
\end{array}\right.
\end{equation}
In particular, the matrix $L_k(\lambda)$ remains isospectral in the discrete 
time evolution described by the equations {\rm(\ref{dTQ})}.
\end{theorem}
More precisely, the first equation in (\ref{dTQ}) is equivalent to the
first equation in (\ref{dTQ fact}) (under the assumption that $W_k\in G=
{\rm SO}(n)$), while the second and the third equations in (\ref{dTQ}) are
equivalent to the second equation in (\ref{dTQ fact}). So, the problem of
solvability of the first equation in (\ref{dTQ}) for $W_k$ is equivalent to
the matrix factorization problem expressed by the first equation in 
(\ref{dTQ fact}). For a treatment of a closely related factorization problem, 
we refer the reader to [MV].

From Theorem 4 there follows also the complete integrability of our
discrete time Lagrangian map. Notice that its integrals of motion {\it do not}
coincide with the integrals of motion of the continuous--time problem (with
the only exception of the free rigid body motion considered in [MV]).
To be more concrete, the integrals of motion of our map are obtained from
the integrals of the continuous time problem by replacing $P$ through
$P-\frac{1}{4}\epsilon^2P^2$. As for the actual integration of our map
in terms of theta--functions, we leave it as an open problem for want of 
a better occasion (cf. [MV], [B], [RSTS]).

\setcounter{equation}{0}
\section{A particular case: the Clebsch problem}

An important particular case of the rigid body in a quadratic potential
appears when $A=aa^{\rm T}$ with some $a\in{\mathbb R}^n$. For example, this
is the case when the quadratic potential (\ref{TQ pot}) represents the
quadratic terms in an expansion of a potential of a single point mass;
it is supposed that the distance from the rigid body to this point mass is much
larger than the size of the body itself, and the ratio of this two length
scales is the small parameter of the above mentioned expansion. In this case
the vector $a$ points from the point mass to the fixed center of mass of the
rigid body. Then
\begin{equation}\label{TQ Clebsch p}
p=g^{\rm T}a
\end{equation}
represents the same vector in the frame firmly attached to the rigid body.
We have:
\begin{equation}\label{TQ Clebsch P}
P=pp^{\rm T}\;,
\end{equation}
i.e. the orbit $O_A$ in $V$ consists of rank 1 matrices. This case could be
considered independently, along the lines of this paper. The relevant 
representation space $V$ would be then ${\mathbb R}^n$, and the representation
$\Phi$ of $G$ in $V$ would be defined as $\Phi(g)\cdot v=gv$. However, we
prefere to simply use the results already obtained, replacing $A$ with
$aa^{\rm T}$, and $P$ with $pp^{\rm T}$. So, the Lagrange function
(\ref{TQ lagr 2}) takes the form
\begin{equation}\label{TQ Clebsch Lagr}
\bL(g,\dot{g})=\cL^{(l)}(\Omega,p)=
-\frac{1}{2}\,{\rm tr}(\Omega J\Omega)+\frac{1}{2}\,\langle p,Jp\rangle\;,
\end{equation}
where now $\langle\cdot,\cdot\rangle$ denote the standard scalar product in
${\mathbb R}^n$. The equations of motion (\ref{TQ}) become
\begin{equation}\label{TQ Clebsch}
\left\{\begin{array}{l} \dot{M}=[M,\Omega]+p\wedge (Jp)\;,\\ \\ 
\dot{p}=-\Omega p\;,\end{array}\right.
\end{equation}
where the notation $p\wedge q=pq^{\rm T}-qp^{\rm T}$ for $p,q\in {\mathbb R}^n$
is used. This system is Hamiltonian on the dual of the semidirect product Lie
algebra ${\rm e}(n)={\rm so}(n)\ltimes {\mathbb R}^n$, with the Hamilton
function
\begin{equation}\label{TQ Clebsch Ham}
H(M,p)=\frac{1}{2}\langle M,\Omega\rangle-\frac{1}{2}\,\langle p,Jp\rangle=
\frac{1}{2}\sum_{j<k} \frac{M_{jk}^2}{J_j+J_k}-\frac{1}{2}\sum_{k=1}^n
J_kp_k^2\;.
\end{equation}
The integrable discretization of this system is given by the discrete time
Lagrange function
\begin{equation}\label{dTQ Clebsch Lagr}
\bbL(g_k,g_{k+1})=\Lambda^{(l)}(W_k,p_k)=
-\frac{1}{\epsilon}\,{\rm tr}(W_kJ)
+\frac{\epsilon}{2}\,\langle p_{k+1},Jp_k\rangle=
-\frac{1}{\epsilon}\,{\rm tr}(W_kJ)
+\frac{\epsilon}{2}\,\langle p_k,W_kJp_k\rangle\;,
\end{equation}
where, as usual, $W_k=g_k^{\rm T}g_{k+1}\in G$ and $p_k=g_k^{\rm T}a\in O_a$.
The equations of motion of this discretization read: 
\begin{equation}\label{dTQ Clebsch}
\left\{\begin{array}{l}
M_k=\displaystyle\frac{1}{\epsilon}(W_kJ-JW_k^{\rm T})-
    \displaystyle\frac{\epsilon}{2}\,p_k\wedge (Jp_{k+1})\;\\ \\
M_{k+1}=\displaystyle\frac{1}{\epsilon}(JW_k-W_k^{\rm T}J)+
        \displaystyle\frac{\epsilon}{2}\,p_{k+1}\wedge (Jp_k)\;,\\ \\
p_{k+1}=W_k^{\rm T}p_k\;.
\end{array}\right.
\end{equation}
The Lax matrix and the Lax representation of this map are obtained from
(\ref{dL}) and Theorem 5 by replacing $P_k$ through $p_kp_k^{\rm T}$.

It remains to be noticed that originally the integrable Hamiltonian
(\ref{TQ Clebsch Ham}) was found (in the ``physical'' case $n=3$) by
Clebsch [C] in another setting, namely in the problem of a motion of a rigid
body in an ideal fluid. (An $n$--dimensional generalization is due to 
Perelomov [P]). In this setting the system appears as a result of reduction of
a Lagrangian on the group ${\rm E}(n)$, left--invariant under the action of
a whole group, rather than a Lagrangian on ${\rm SO}(n)$, left--invariant
under the action of an isotropy subgroup of $a$. Nevertheless, these two
different settings lead to formally identical results.

\section{Conclusion}
The model introduced in the present paper serves as a further important
example of the completely integrable Lagrangian systems with a discrete time
\`a la Moser--Veselov. Actually, this is the second example (after [BS1])
where the version of the discrete Lagrangian reduction is essential, leading
to systems on duals to semidirect product Lie algebras. This new application
was made possible due to the theoretical development in [BS2]. Probably,
the type of Lagrangians introduced in these papers is able to produce
further interesting examples, important for applications. At this point,
I would like to express my gratitude to A.Bobenko, the collaboration with
whom in [BS1], [BS2] was crucial also for this work.

Generally, we consider the discrete time Lagrangian mechanics as an important 
source of symplectic and, more general, Poisson maps. From some points of view 
the variational (Lagrangian) structure is even more fundamental and important 
than the Poisson (Hamiltonian) one (cf. \cite{HMR}, \cite{MPS}, where a similar 
viewpoint is represented). A very intriguing and still not completely 
understood point is a capability of the discrete Lagrangian approach to 
produce completely integrable systems. It would be highly desirable to
continue the search for integrable Lagrangian discretizations of the 
known integrable systems. Hopefully, the list of such discretizations
established in [V],[MV], [BS1], and the present paper, will be further 
extended. Also the generalizations to the infinite dimensional case, 
e.g. to discretization of ideal compressible fluids motion (see [HMR]), 
are highly desirable.


\end{document}